\pdfoutput=1
\documentclass[%
reprint,
amsmath,amssymb,
aps,
prl,
floatfix,
]{revtex4-2}

\usepackage[pdftex]{graphicx}
\usepackage{dcolumn}
\usepackage{bm}

\newcommand{\simge}{\;\lower3pt\hbox{$\stackrel{\textstyle >}{\sim}$}\;}
\newcommand{\simle}{\;\lower3pt\hbox{$\stackrel{\textstyle <}{\sim}$}\;}
\newcommand{\deltaphi}{\left| {\Delta} \phi_x \right|}

\begin{document}


\title{Critical Behavior at the Onset of Multichimera States in a  Coupled-Oscillator  Array}
\author{Katsuya Kawase}
\author{Nariya Uchida} \email{uchida@cmpt.phys.tohoku.ac.jp}
\affiliation{Department of Physics, Tohoku University, Sendai 980-8578, Japan}
\date{\today}
\begin{abstract}
We  numerically investigate the onset of multi-chimera states in a linear array
of coupled oscillators. 
As the phase delay $\alpha$ is increased, they exhibit a continuous transition from 
the globally synchronized state to the multichimera state consisting of
asynchronous and synchronous domains. 
Large-scale simulations show that
the fraction of asynchronous sites $\rho_a$ obeys the power law  
$\rho_a \sim (\alpha - \alpha_c)^{\beta_a}$,
and that the spatio-temporal gaps between asynchronous sites show power-law distributions
at the critical point.
The critical exponents are distinct from 
those of the (1+1)-dimensional directed percolation
and other absorbing-state phase transitions, 
indicating that this transition belongs to a new class of non-equilibrium critical phenomena.
Crucial roles are played by traveling waves that rejuvenate 
asynchronous clusters by mediating non-local interactions between them.
\end{abstract}

\maketitle


 
Synchronization is a ubiquitous phenomenon in Nature
and also is of vital importance for our life, as seen in 
the coordinated contraction of cardiac cells and firing of neurons.  
The mode of synchronization crucially depends on the range of  interation between oscillators.
Collective synchronization is realized by global and power-law couplings, 
while finite-range coupling lead to  local and partial coherence.
A surprising phenomenon in the latter is the spatial coexistence of 
coherent and incoherent clusters, called {chimera states}~\cite{abrams2004chimera}.
Since its discovery by Kuramoto and Battogtokh~\cite{kuramoto2002coexistence},
chimera states have been a target of intense research activities~\cite{abrams2004chimera,shima2004rotating,
kim2004pattern,abrams2006chimera,
abrams2008solvable,omel2008chimera,bordyugov2010self,
omel2013coherence,sethia2014chimera,yeldesbay2014chimeralike,
panaggio2015chimera,
laing2009dynamics,laing2009chimera,
shanahan2010metastable,omelchenko2015robustness,scholl2016synchronization,
hizanidis2016chimera,majhi2016chimera,
tinsley2012chimera,nkomo2013chimera,totz2018spiral,
martens2013chimera,kapitaniak2014imperfect,
rosin2014transient,gambuzza2014experimental,
bohm2015amplitude,
wolfrum2011chimera,
sethia2008clustered,sethia2013amplitude,omel2013nonlocal,gopal2014observation,
xie2014multicluster,omelchenko2015nonlinearity,xie2015twisted,suda2018breathing,
omel2014partially,wolfrum2016turbulence,omel2018mathematics}.
What makes them mysterious  is the emergence of heterogeneity
from identical oscillators with uniform network topology, while they are also 
obtained in heterogeneous networks~\cite{laing2009dynamics,laing2009chimera,
shanahan2010metastable,omelchenko2015robustness,scholl2016synchronization,
hizanidis2016chimera,majhi2016chimera}.
An essential ingredient in generating chimera states is 
the frustrated coupling with phase delay.  
The onset of chimera states and bifurcation scenarios 
have been studied using a variety of models~\cite{kim2004pattern,abrams2006chimera,
abrams2008solvable,omel2008chimera,bordyugov2010self,
omel2013coherence,sethia2014chimera,yeldesbay2014chimeralike}.
Chimera states are experimentally reproduced by 
electro-chemical~\cite{tinsley2012chimera,nkomo2013chimera,totz2018spiral},
mechanical~\cite{martens2013chimera,kapitaniak2014imperfect}, 
electronic~\cite{rosin2014transient,gambuzza2014experimental} and 
optical~\cite{bohm2015amplitude} oscillators.
When the number of oscillators $N$ is large compared to the coupling range $L$,
multiple clusters of coherent and incoherent domains appear 
(multichimera states)~\cite{sethia2008clustered,sethia2013amplitude,omel2013nonlocal,gopal2014observation,
xie2014multicluster,omelchenko2015nonlinearity,xie2015twisted,suda2018breathing}.
In a fully developed multichimera state,
the number of clusters roughly scales as $m \sim N/L$, and phase diagrams 
have been obtained for transition between different numbers of clusters.
A continuum limit approach 
was taken to analyze the stability of traveling wave solutions and 
their transition to turbulence~\cite{omel2014partially,wolfrum2016turbulence,omel2018mathematics}.
However,  the onset and statistical properties of multichimera states
with $m \gg 1$ are still largely unexplored,
which are the subject of the present paper.
Near the onset, we find branching and self-proliferating patterns of clusters similar to those 
found in directed percolation (DP)~\cite{hinrichsen2000non} 
and observed at the onset of various turbulence phases~\cite{takeuchi2007directed,lemoult2016directed,sano2016universal}.
Meanwhile, 
the fraction of asynchronous sites as well as the gap between clusters
show critical behaviors that are distinct from those of DP and other 
absorbing-state phase transitions,
indicating that the onset of multichimera states belongs 
to a new class of non-equilibrium critical phenomena.
We demonstrate that the characteristic spatio-temporal properties are caused by
traveling waves that mediate non-local interaction between asynchronous clusters.


The chimera states are obtained for a variety of coupling functions.
We employ a step function with the coupling range $L$.
Assuming that the intrinsic frequencies are the same for all oscillators,
which are set to zero without losing generality, 
we use the model equation 
\begin{equation} 
\label{eq:model}
 \dot{\phi}_x = - \frac{1}{2L} \sum_{0 < |s| < L}
 \sin \left(  \phi_x - \phi_{x+s} + \alpha \pi \right),
\end{equation}
where $\phi_x = \phi_x(t)$ is the phase at the integer coordinate $x$ and time $t$, 
and $\alpha (>0)$ gives the phase delay.
Each pair of oscillators  tend to synchronize in-phase for $\alpha< \frac12$,  
and anti-phase for $\frac12 < \alpha < 1$.
Frustration due to the non-local coupling 
destroys uniform synchronization  and gives rise 
to the chimera states for $\alpha$ near $\frac12$.
We numerically solved (\ref{eq:model}) with
a periodic boundary condition
for the system size $N$ up to $2^{24}=16777216$. 
The coupling range $L=5$ is used unless otherwise stated.
By choosing $L \ll N$,  we can study the statistical behavior of 
a large number of asynchronous clusters.
A uniformly random and spatially uncorrelated distributions of $\phi_x$ in $[0,2\pi]$ is used for the initial condition.
In order to quantify the degree of local synchronization, 
we define the reduced phase difference 
\begin{equation}
\deltaphi =  \left| \left[ \frac{\phi_{x+1} - \phi_x}{\pi} \right] \right| \in [0,1],
\end{equation}
where $[p] \equiv p -  2 \lfloor \frac{p+1}{2} \rfloor$ means truncation into $[-1,1]$.

\begin{figure*}[h]
\centering
\includegraphics[bb=0.000000 0.000000 4496.000000 2242.000000,width=175mm]{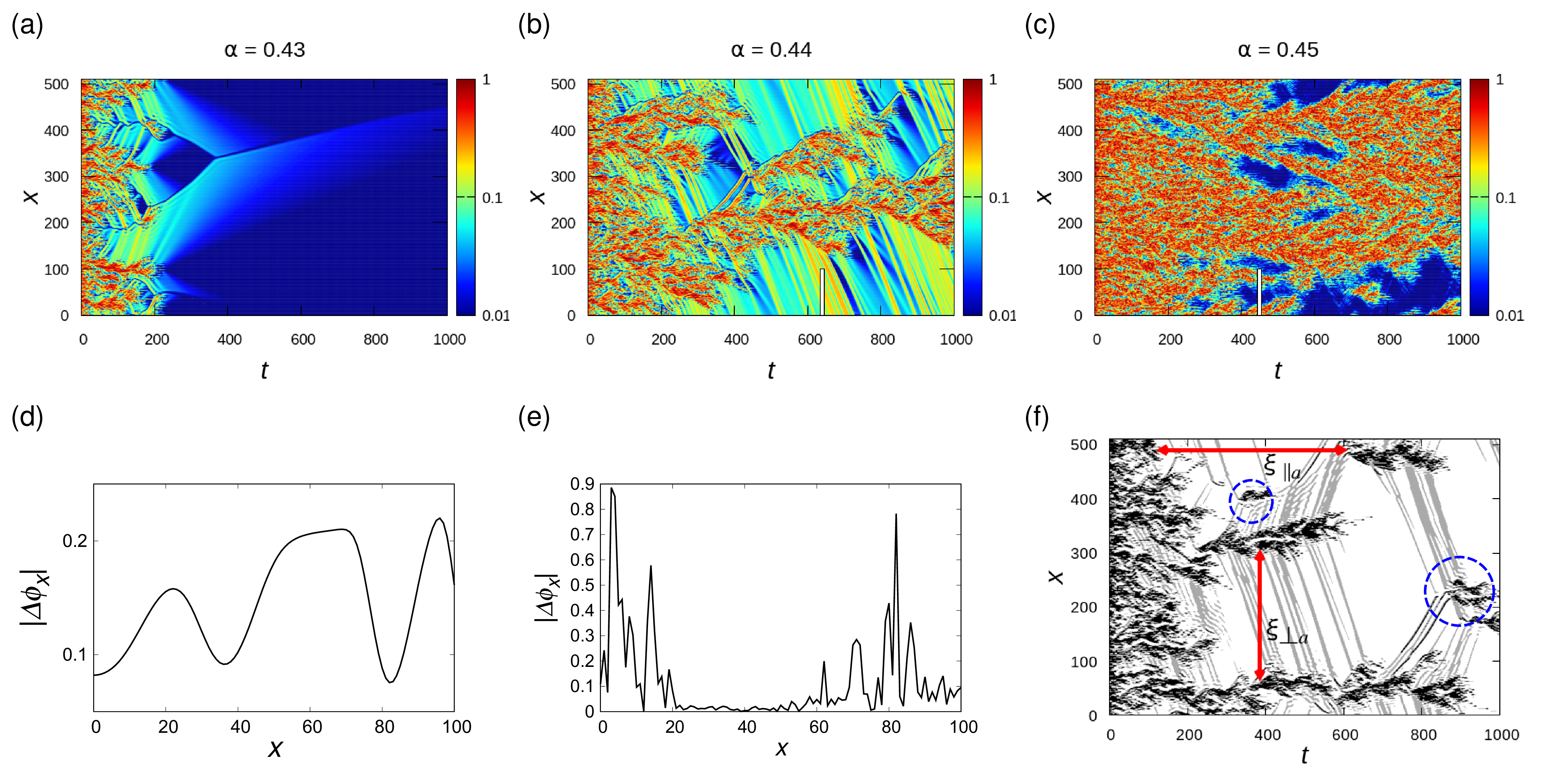} 
\caption{\label{fig:colormap}
Spatio-temporal patterns of the phase difference for 
(a) $\alpha=0.43$, (b) $\alpha=0.44$, and (c) $\alpha=0.45$ in an $N=512$ system.
They are dominated by synchronized, traveling waves and multichimera states
in the late stage, respectively. 
The spatial profiles of $\deltaphi$ in (d) and (e) are 
taken at the time indicated by the white bars in (b) and (c), respectively.
The smooth wavy profile with $\deltaphi \simeq 0.1$ in (d) is characteristic to the traveling wave states.
In (e), we see coexistence of multichimera and synchronized domains. 
A multichimera domain consists of asynchronous sites ($\deltaphi > \Delta_2$) 
and synchronized sites ($\deltaphi < \Delta_1$).  
(f) spatio-temporal plot of the ternarized phase difference for $\alpha=0.44$. 
The asynchronous sites are shown in black, the traveling-wave sites in grey, and the synchronous sites in white. 
The spatial (temporal) gap $\xi_{\perp a}$ ($\xi_{\parallel a}$) between asynchronous sites are illustrated by arrows.
Collision of traveling waves generate new asynchronous clusters in the regions enclosed by dotted circles.
}
\end{figure*}

In Fig.~\ref{fig:colormap}, we show the spatio-temporal map of the phase difference 
for different values of $\alpha$. 
For $\alpha = 0.43$, the system reaches a uniform synchronized state with $\deltaphi \ll 1$ 
at large $t$. 
For  $\alpha=0.45$, we observe a multichimera state with densely packed asynchronous clusters 
and a small fraction of synchronized domains.  
For the intermediate value $\alpha = 0.44$, we find a branching and self-proliferating structure 
of asynchronous clusters that emit a number of traveling waves with $\deltaphi \simeq 0.1$.
The number fraction of asynchronous clusters slowly decays in time
and the traveling waves become dominant in the late stage.
Spatial profiles of $\deltaphi$ in Fig.\ref{fig:colormap} (d)(e) show 
a marked contrast between the traveling waves, multichimera and synchronized states.
While traveling wave states have a smooth profile with a characteristic wavelength $\lambda \gg L$,
the multichimera states have a fluctuating noisy profile with a typical size $\sim L$.

The branching structure of asynchronous clusters 
reminds us of the similar structure in DP.
In (1+1)-dimensional DP, each bond is created at probability $p$ 
between an active site $(x,t)$ and its neighbors $(x\pm \frac12, t+1)$. 
A site becomes active if it is bonded from one or more neighbors, 
and inactive otherwise.   
As $p$ is varied,
the fraction $\rho$ of the active site at time $t\to\infty$ shows the critical behavior
$\rho \sim (p-p_c)^\beta$. 
At the critical point $p=p_c$, the active fraction 
decays in time as $\rho \sim t^{-\alpha}$, and 
the distribution of spatial (temporal) gap $\xi_\perp$ ($\xi_\parallel$) between active sites 
shows the power-law decay $n(\xi_{i}) \sim \xi_{i}^{-\mu_i}$ $(i = \parallel,\perp)$.
We analyze the transition from the synchronized to multi-chimera states
in light of the critical behaviors of DP.

\begin{figure}[htb]
\centering
\includegraphics[width=86mm]{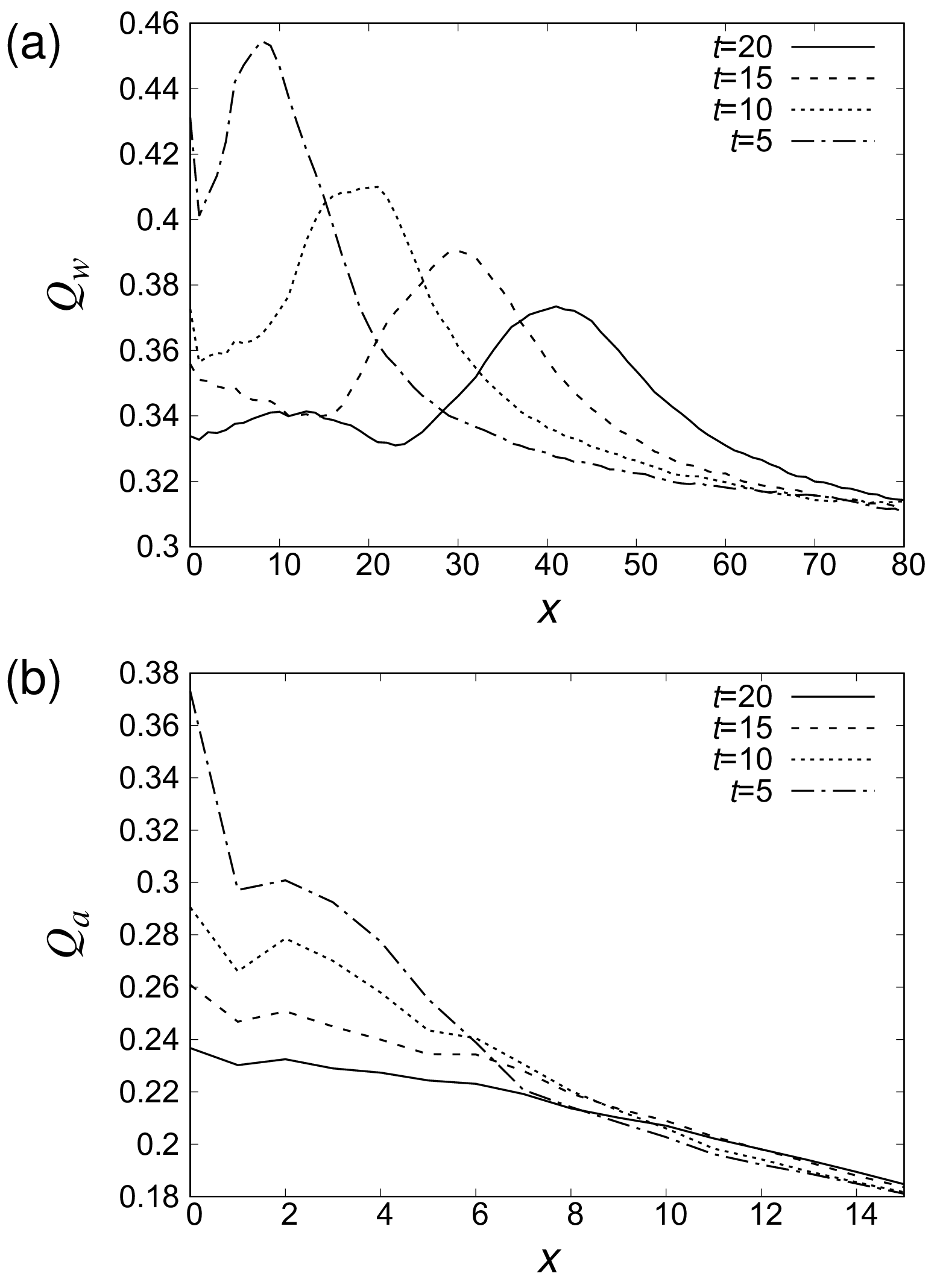}
\vspace{70mm}
\caption{\label{fig:QA}
Spatial profiles of the correlation functions 
(a) $Q_w(x,t)$ and (b)  $Q_a(x,t)$ 
for different time delay $t$.
The average is taken over the time interval $250<t'<1000$ for an $N=131072$ system.
The correlation function of traveling wave sites has a peak that moves with a speed $v_w \simeq 2$,
while that of asynchronous sites has a peak at $x=0$ and a shoulder in the region $x \simle L=5$.
}
\end{figure}

%
%
To this end, it is necessary to define the asynchronous and synchronous sites, 
which correspond to the active and inactive sites in DP, respectively.
The traveling wave states are characteristic to our system, 
and  they produce a crucial difference from DP, as we shall see below.
Therefore, we classify the sites into the three types: 
($s$) synchronized, ($w$) traveling waves, and ($a$) asynchronous.
We find that the phase difference $|{\Delta} \phi_x|$ is a particularly 
good measure to distinguish them, when compared to other measures such as 
the phase velocity $\dot{\phi}_x$ 
and 
the order parameter defined by $R_x = \frac{1}{2L+1} 
\left|\sum_{|s| <= L} e^{i \phi_{x+s}} \right|$ . 
We use the criteria
\begin{eqnarray}
(s) \quad & \deltaphi \in [0,\Delta_1], 
\\
(w) \quad & \quad \deltaphi \in [\Delta_1, \Delta_2],
\\
(a) \quad & \deltaphi \in [\Delta_2, 1]
\end{eqnarray}
with the thresholds $\Delta_1=0.1$ and $\Delta_2 = 0.3$.
With this choice of $\Delta_1$ and $\Delta_2$, 
we can distinguish the peaks and valleys of the traveling waves, 
which belong to $(w)$ and $(s)$, respectively. 
A spatio-temporal map of the ternarized phase difference is 
shown in Fig.\ref{fig:colormap} (f).
It clearly differentiates the branching asynchronous clusters 
and the traveling waves. 
Using the state variable $\sigma_{x,t} = s, w, a$,
we define the spatio-temporal correlation function 
$Q_i(x,t)$ as the average of the conditional probability that 
$\sigma_{x+x', t+t'}= i$ under the condition $\sigma_{x',t'} = i$:
\begin{equation}
Q_i(x,t) = \langle \, {\rm Prob}(\sigma_{x+x',t+t'}=i | \sigma_{x',t'} =i ) \, \rangle_{x',t'},  
\end{equation}
where $i = s,w,a$. In Fig.\ref{fig:QA}, we show 
the correlation functions $Q_w(x,t)$ and $Q_a(x,t)$ for $\alpha = 0.44$. 
As a function of the distance, the former has a moving peak with velocity $v_w \simeq 2$,
indicating wave propagation.
On the other hand, the asynchronous sites have only a short-range correlation
with a peak at $x=0$ and a shoulder in the coupling range $L$. 
The choice of $\Delta_1$ and $\Delta_2$ is made to maximize the difference 
between the two correlation patterns; see Supplementary Material (SM) for details.

\begin{figure}[htb]
\centering
\includegraphics[bb=0.000000 0.000000 1604.000000 2067.000000,width=86mm]{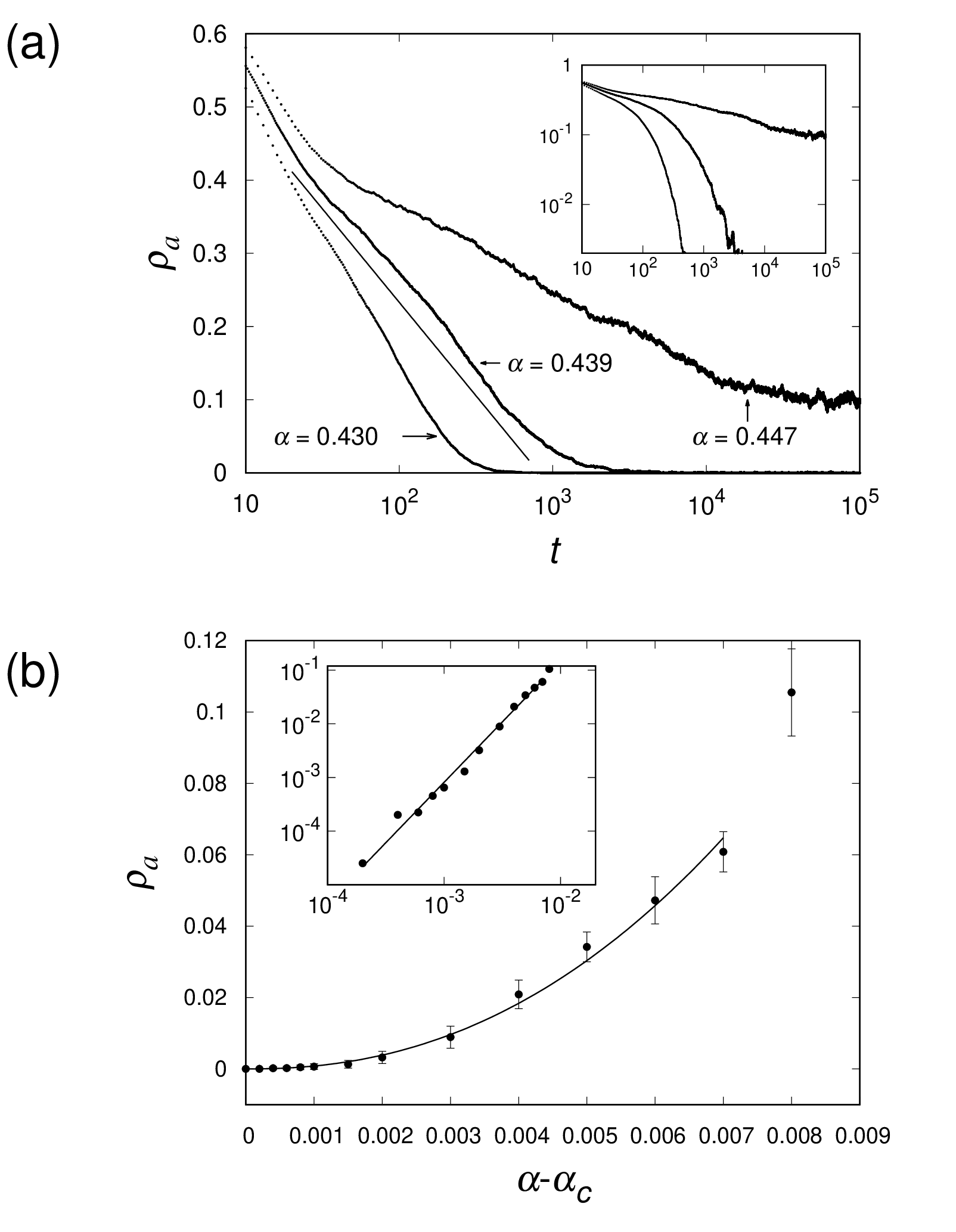}
\caption{\label{fig:rho_a}
(a) The fraction of asynchronous sites $\rho_a(t)$ for an $N=131072$ system.
It decays to zero for  $\alpha\le \alpha_c= 0.4390$.
(b) The steady state fraction $\rho_a(t\to\infty)$ versus  $\alpha-\alpha_c$.
Inset: logarithmic plot. 
Solid lines show the power-law fitting with the exponent $\beta_a = 2.21$.
}
\end{figure}

Now we analyze the transition behavior in terms of the fraction of the asynchronous sites $\rho_a(t)$, 
which is plotted in Fig.\ref{fig:rho_a}.
Using the system size $N=131072$, we find that $\rho_a$ decays to zero
after a long time for $\alpha \le \alpha_c = 0.4390$,
while it remains non-zero for $\alpha > \alpha_c$.
At the critical point $\alpha_c$, the initial decay is fitted by $\rho_a(t) \sim -\ln(t/t_1)$ 
with the characteristic time $t_1 =1.1\times 10^3$.
The logarithmic decay crossovers to a slower decay at $t \simeq t_1$. 
At $t=t_1$, the asynchronous fraction is already as small as  0.025, 
but it vanishes only after $t=t_0 \simeq 5\times  10^5 \simeq 5\times 10^2 \times t_1$.
During this late stage, $\rho_a(t)$  shows a noisy profile with many peaks
due to traveling waves, as discussed later and in SM.

The asynchronous fraction $\rho_a$ in the dynamical steady state $t \to \infty$
is estimated by averaging over the time window $10^5 < t < 10^6$
except for a few data points near the critical point, 
for which the time window is shifted 
to minimize the statistical error.
We plot the steady state fraction versus $\alpha - \alpha_c$ in Fig.~\ref{fig:rho_a} (b).
It is fitted by the  power law 
\begin{equation}
\rho_a \sim (\alpha-\alpha_c)^{\beta_a}, \quad \beta_a = 2.21 \pm 0.09.
\end{equation}
The exponent is much larger than that of the fraction of active sites in (1+1)-dimensional DP,
$\beta_{\rm DP}\simeq 0.277$.

\begin{figure}[htb]
\centering
\includegraphics[bb=0.000000 0.000000 1508.000000 2112.000000,width=86mm]{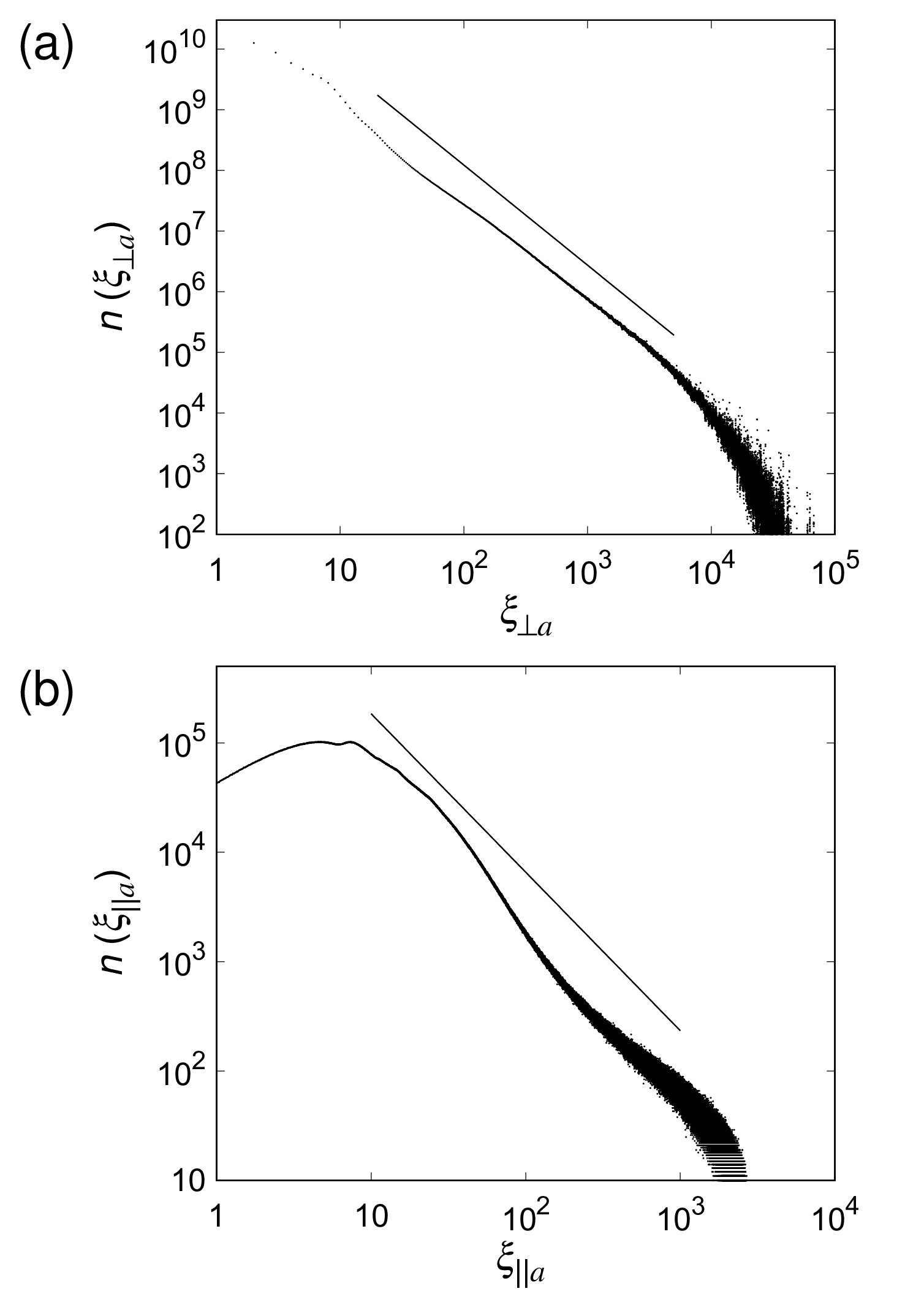} 
\caption{\label{fig:kankaku}
Histograms of (a) the spatial gap $\xi_{\perp a}$ and
(b) temporal gap $\xi_{\parallel a}$ between asynchronous sites for $\alpha = \alpha_c$.
Solid lines show the power-law fitting with the exponents 
$\mu_{\perp a} = 1.71$ and $\mu_{\parallel a} = 1.73$, respectively.
}
\end{figure}

Next, we consider the spatial gap $\xi_{\perp a}$ and the temporal gap $\xi_{\parallel a}$ 
between asynchronous sites,  illustrated by the arrows in Fig.~\ref{fig:colormap} (f).
They are measured over the time window $0 < t <3000$ in an $N=16777216$ system.
The histograms of the spatio-temporal gaps at $\alpha = \alpha_c$ are shown 
in Fig.~\ref{fig:kankaku}.
We fitted them by the power law 
\begin{equation}
n(\xi_{\perp a})  \sim \xi^{- \mu_{\perp a}}, 
\quad 
n(\xi_{\parallel a})  \sim \xi^{- \mu_{\parallel a}}
\end{equation}
and estimated  the exponents as 
\begin{equation}
\mu_{\perp a}  = 1.71 \pm 0.01, 
\quad
\mu_{\parallel a}  = 1.73 \pm 0.01. 
\end{equation}
These exponents are smaller than 
those of the active sites in (1+1)-dimensional DP, 
$\mu_{\perp {\rm DP}} \simeq 1.748$ 
and 
$\mu_{\parallel {\rm DP}} \simeq 1.841$.

\begin{figure}[htbp]
\centering
\includegraphics[bb=0.000000 0.000000 1554.000000 2112.000000,width=86mm]{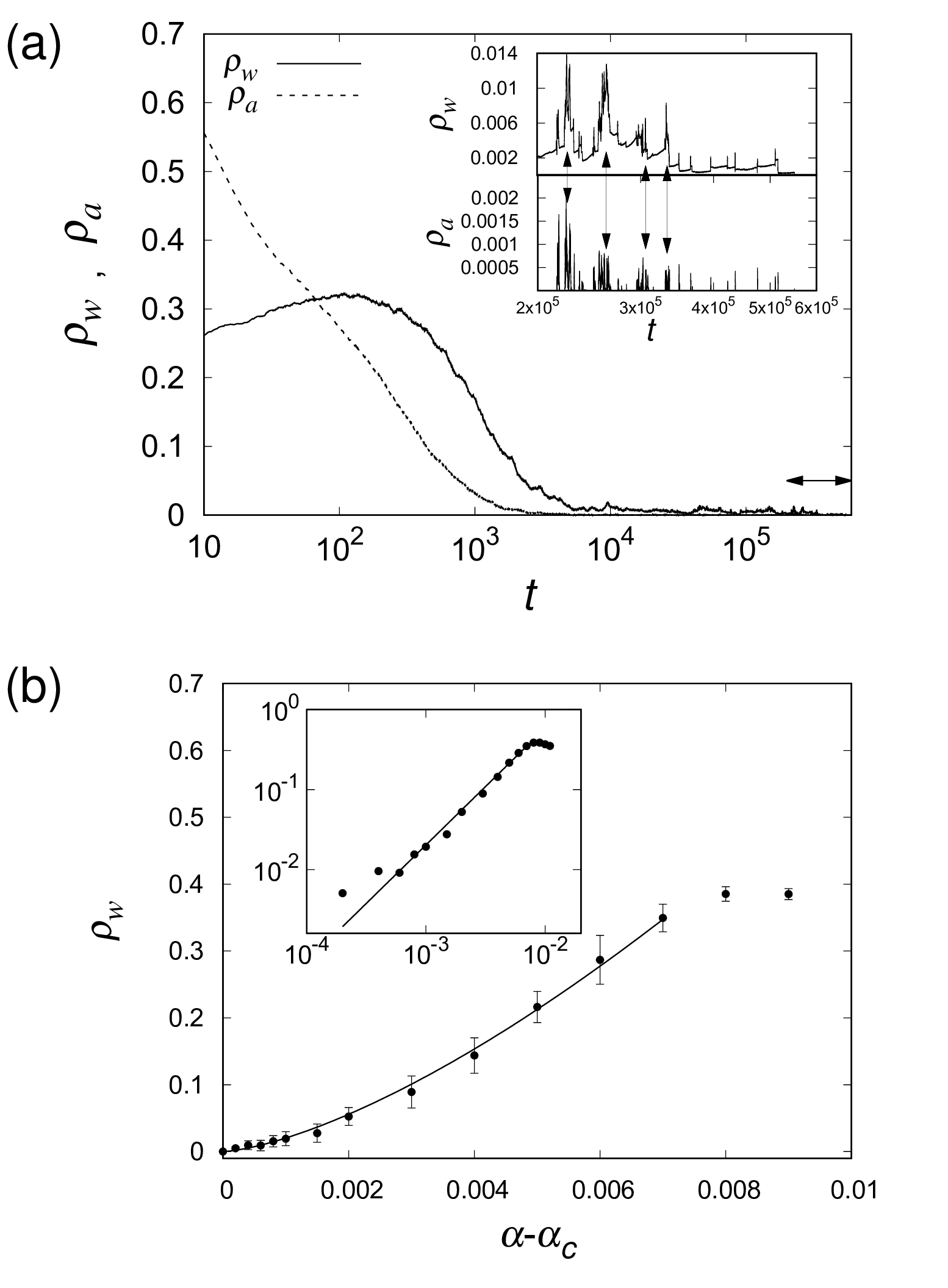} 
\caption{\label{fig:rho_w}
(a) Time evolution of the fraction of traveling wave sites $\rho_w(t)$ 
and asynchronous sites $\rho_a(t)$ for $\alpha = \alpha_c$.
Inset:  enlarged view of the time window $2\times 10^5 < t< 6\times 10^5$.
(b) Steady state fraction of traveling wave sites $\rho_w(t\to \infty)$.
Inset: logarithmic plot.
Solid lines show the power-law fitting with the exponent $\beta_w = 1.46$.
}
\end{figure}

We seek the origin of the novel critical behaviors 
in the traveling waves that are absent in DP. 
We find that the asynchronous states are rejuvenated by collision 
of traveling waves propagating in the opposite directions,
as shown in the dotted circles in Fig.~\ref{fig:colormap} (f).
Time evolution of the fraction of traveling wave sites $\rho_w$ 
at $\alpha = \alpha_c$ is shown in Fig.~\ref{fig:rho_w}(a). 
After an initial growth up to $t\simeq 200$,  it decays rapidly until $t\simeq 5000$. 
In the late stage, $\rho_w$ slowly decays to zero exhibiting many spikes 
that are in correlation with those of the asynchronous sites (see the inset).
These spikes are caused by a positive feedback loop consisting of 
wave emission by asynchronous sites and 
rejuvenation of asynchronous sites by colliding waves. 
We plot the fraction of traveling wave sites $\rho_w$ 
in the dynamical steady states versus $\alpha$ in Fig.~\ref{fig:rho_w}(b). 
It is fitted by the power law
\begin{equation}
\rho_w \sim (\alpha - \alpha_c)^{\beta_w} , \quad \beta_w = 1.46 \pm 0.06.
\end{equation}
The exponent is smaller than that of the asynchronous fraction, which means 
that more traveling waves are needed to rejuvenate an asynchronous site 
as we approach the critical point.


The critical properties of the asynchronous fraction are not explained by  DP 
and related models with local interactions, such as compact DP~\cite{essam1989directed}
and the Ziff-Gulari-Barshad model~\cite{ziff1986kinetic}
which allows rejuvenation by catalytic reactions.
The anomalous DP with non-local spreading rules~\cite{mollison1977spatial,hinrichsen2000non}
is not applicable to  our case, either.
Therefore, we conclude that the onset of  multichimera states
belongs to a new class of non-equilibrium critical phenomena. 
The traveling waves mediate non-local interactions between asynchronous clusters
and rejuvenate them, leading to the slow decay of the asynchronous fraction. 
The precise  mechanism of the non-local interactions and its relation to 
the critical behavior are important questions to be addressed in future work.



%
%
%
%
%
%
%
%
%
%
%

\end{document}